\documentclass[9pt,singlecolumn,twoside]{opticajnl}
\journal{opticajournal} % use for journal or Optica Open submissions

\setboolean{shortarticle}{False}
\usepackage{lineno}
\usepackage{graphicx}% Include figure files
\usepackage{dcolumn}% Align table columns on decimal point
\usepackage{bm}% bold math
\usepackage{mathtools}
\usepackage{siunitx}
\usepackage{upgreek}
\usepackage{babel}
\usepackage{url}

\title{Fabrication-Aware Inverse Design of Nanophotonic Devices for 3D Laser-Nanoprinting}

\author[1*]{Oliver Kuster}
\author[2]{Tim Alletzhäusser}
\author[1, 3, 4]{Carsten Rockstuhl}
\author[2, 3]{Martin Wegener}
\author[3]{Thomas Jebb Sturges}

\affil[1]{Institute of Theoretical Solid State Physics, Karlsruhe Institute of Technology, Kaiserstrasse 12, 76131, Karlsruhe, German}
\affil[2]{Institute of Applied Physics, Karlsruhe Institute of Technology, Kaiserstrasse 12, 76131, Karlsruhe, Germany}
\affil[3]{Institute of Nanotechnology, Karlsruhe Institute of Technology, Kaiserstrasse 12, 76131, Karlsruhe, Germany}
\affil[4]{Center for Integrated Quantum Science and Technology (IQST), Karlsruhe Institute of Technology, Kaiserstrasse 12, Karlsruhe, 76131, Germany}

\affil[*]{oliver.kuster@kit.edu}

\begin{abstract}
Advances in 3D laser-nanoprinting enable us to fabricate 3D nanophotonic devices with a wide range of functionalities on demand.
By exploiting all three spatial dimensions, an enormous design space becomes available for these nanophotonic devices. 
However, such an immense design space is impossible to explore efficiently by intuition alone, especially when designing free-form nanophotonic devices.
Density-based topology optimization offers a natural tool for 3D nanophotonic design by allowing the efficient design of devices with millions of degrees of freedom.
Traditional density-based topology optimization relies on heuristic measures to account for limitations imposed by the fabrication method.
Indeed, the fabrication method is rarely considered as part of the forward model in the design pipeline.
In this work, we introduce an inverse design method that explicitly models the direct-laser-writing process used in 3D nanoprinting. 
Incorporating a differentiable formulation of the direct-laser-writing model allows us to design 3D nanophotonic devices within the experimentally available design space and to precompensate for fabrication-specific effects. Optimizing inside the experimentally available design space ensures that the constraints we put on the optimization are given by our parametrization of the fabrication method and not by heuristic methods, which might over- or underconstrain the optimization problem.
Furthermore, modeling the 3D laser-nanoprinting process explicitly allows us to not only take fabrication-specific effects, such as the proximity effect, into account but also enables the optimization to actively make use of these fabrication-specific effects to increase the functionality of the device.
\end{abstract}

\setboolean{displaycopyright}{false} % Do not include copyright or licensing information in submission.

\begin{document}

\maketitle

\section{\label{sec:introduction}Introduction}

\begin{figure*}[t]
    \centering
    \includegraphics[width=0.9\linewidth]{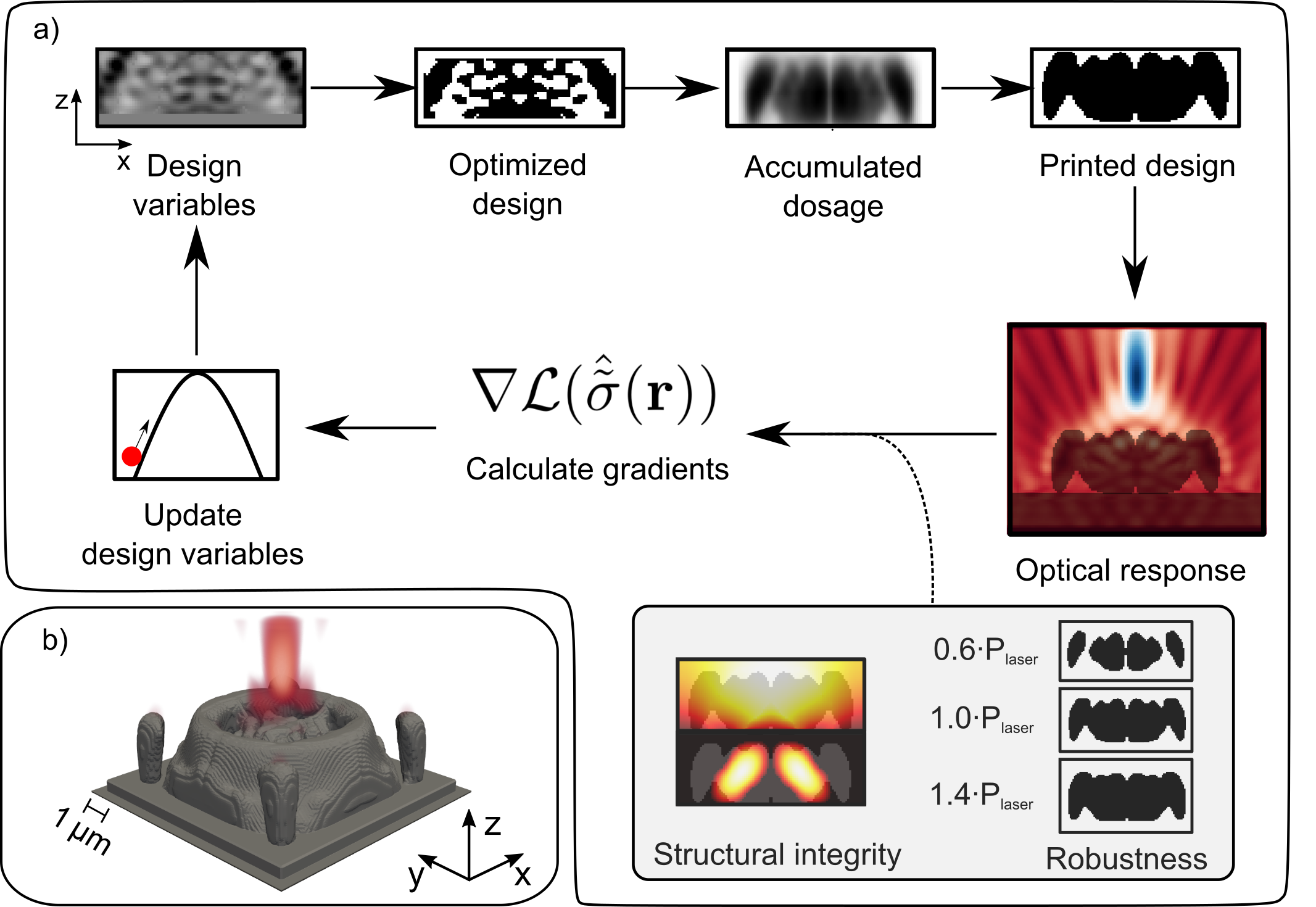}
    \caption{a) Schematic of the design workflow of the DLW-Model at the example of a metalens. Please note that we sketch all quantities here in a 2D cross-sectional view, but the entire analysis is done in 3D. Starting from the top left, the design variable, \textit{i.e.}, the density, is first converted into a binarized design representing the writing pattern. It is that writing pattern that specifies at which voxel the laser is switched on to write the structure in the 3D laser-nanoprinting process.
    This writing pattern is then convolved with the PSF of the laser focus to calculate the accumulated dose in the photoresist. The printed design is then retrieved by identifying the spatial domain where the accumulated dose exceeds the polymerization threshold as material, and as a void otherwise. We then calculate the optical response of the design and incorporate structural integrity (using the virtual temperature method) and robustness to derive the final figure of merit $\mathcal{L}(\hat{\tilde{\sigma}}(\mathbf{r}))$. By using the adjoint method and automatic differentiation, we compute the gradients $\nabla\mathcal{L}(\hat{\tilde{\sigma}}(\mathbf{r}))$ and use them to update the design via a gradient-based optimizer. We elaborate on each of the steps of the design workflow in Section~\ref{sec:methods}.
    b) Functionality of a small metalens designed with the DLW-Model. Such a fabrication-aware inverse design cycle provides the digital blueprint for structures to be written.
    %We note that the designs shown in the workflow on the right are not the same design as shown on the right, but are only used for illustration purposes.}
    }
    \label{fig:dlw-scheme}
\end{figure*}
Topology optimization is a canonical inverse design approach to identify a structure with a predefined functionality \cite{bendsoeTopologyOptimization2004, sigmundTopologyOptimizationApproaches2013a}. Topology optimization relies on the adjoint formalism to compute the gradients of an objective function with respect to all degrees of freedom that parametrize a structure, followed by a gradient descent to optimize the design. More specifically, we rely on density-based topology optimization below. There, the structure is parameterized by a spatially dependent density, which is translated into an actual material distribution through a sequence of filtering steps. Thanks to the rather convenient formulation, topology optimization is a popular inverse design method used across multiple disciplines \cite{bendsoeMaterialInterpolationSchemes1999, pietropaoliThreedimensionalFluidTopology2019, alexandersenDetailedIntroductionDensitybased2023a, sigmundDesignCompliantMechanisms1997, deatonSurveyStructuralMultidisciplinary2014}.
Nanophotonic design, in particular, benefits from using topology optimization, as efficient designs often require unintuitive sub-wavelength features, which makes it difficult to design nanophotonic devices solving only the forward problem and a limited number of design parameters \cite{moleskyInverseDesignNanophotonics2018, hughesAdjointMethodInverse2018, suInverseDesignDemonstration2018, shangInverseDesignedLithiumNiobate2023, christiansenInverseDesignPhotonics2021, hammondTopologyoptimizedDistributed3d2026, christiansenOrdersMagnitudeReduction2026, gedeonTimeDomainTopologyOptimization2023, nandaInverseDesign3D2026, bezickPearSANMachineLearning2026, romashkinaInverseDesignedSuperchiralHot2025, bergMultiagentReinforcementLearning2026, mahlauGradientinformedBayesianInterior2026, lalau-keralyAdjointShapeOptimization2013, chungHighNAAchromaticMetalenses2020, gertlerManyPhotonicDesign2025}.

While topology optimization allows for a completely free-form design approach, it is usually not possible to directly fabricate structures optimized in this way. To increase fabrication feasibility, restrictions must be imposed on the design space, although these are typically applied heuristically. The purpose, nevertheless, must always be to design the structure only within the subspace of parameters that yields feasible devices. Typical examples for approaches to ensure feasibility are a Gaussian filter to smear out the density to accommodate minimum feature sizes, or erosion and dilation filters to improve robustness against fabrication errors.

In this work, we go beyond such a phenomenological treatment of heuristic restrictions and explicitly consider the fabrication process of 3D laser-nanoprinting as a part of the design pipeline to entirely accommodate fabrication-informed constraints \cite{razaFabricationawareInverseDesign2025, qiaoFabricationawareInverseDesign2025, seoPhysicsGuidedFabricationAwareInverse2026}. Such a comprehensive and explicit description of the fabrication process ensures that the optimized design is printable without unnecessarily limiting the design space.
We call our fabrication-aware inverse design model the Direct-Laser-Writing-Model (DLW-Model). The example workflow of the DLW-Model can be seen in Fig.~\ref{fig:dlw-scheme} a), and a metalens designed using the DLW-Model in Fig.~\ref{fig:dlw-scheme} b).

To accommodate the details of the writing process, we must be fully aware of how it works. In 3D laser-nanoprinting, a writing laser is focused into a negative-tone photoresist to trigger a nonlinear polymerization process within the smallest printable feature, called a voxel. Photoinitiator molecules are excited to a higher energy state by simultaneously absorbing two photons, initiating a chemical reaction and polymerization of the voxel.
By precisely controlling the laser focus writing path, complex 3D structures with sub-wavelength-sized features can be printed.
The smallest printable feature size of a voxel depends on several factors, including the shape of the laser focus, the laser power, and the polymerization threshold of the photoresist.
The final, printed structure can then be obtained by developing the photoresist.

The size of the full-width-half-maximum of the laser focus is typically a few hundred nanometers laterally and several times larger axially. In general, a simple Gaussian or circular filter is not sufficient to capture the shape of the printed voxel. A more accurate representation of the laser focus shape is required. Such a representation is given by calculating the Point-Spread-Function (PSF) \cite{richardsElectromagneticDiffractionOptical1959, sedovaDifferentiableForwardModeling2026}.
Additionally, the dose accumulation in the photoresist plays a crucial role. Polymerization can happen through repeated illumination of the same spot even if each individual illumination would leave the photoresist below the polymerization threshold in an effect known as the proximity effect \cite{barner-kowollik3DLaserMicro2017, chenRecentAdvancesChallenges2025, liu3DLaserNanoprinting2023}.
We want to note, that in principle not only the spatial pattern of the exposure matters, but also the temporal sequence in the so-called spatio-temporal proximity effect, since the path the writing laser takes also influences the final printed structure \cite{wallerSpatioTemporalProximityCharacteristics2016}. However, we only consider the spatial proximity effect in this work.

Such process-specific effects are absent in conventional topology optimization formulations, in which the design variables directly parameterize a nominal geometry whose electromagnetic response is optimized. This nominal geometry is then used as the fabrication target, even though the printing process systematically transforms it into a different fabricated structure. This discrepancy usually means that the optimized designs have to be iteratively improved to bring the designs into the experimentally available design space, leading to a waste of resources and often suboptimal designs \cite{kieferSensitivePhotoresistsRapid2020, liarosMethodsDeterminingEffective2021, howardRelationshipsConversionTemperature2010, yangSchwarzschildEffect3D2019, bauerProgrammableMechanicalProperties2019, shuklaSubwavelengthDirectLaser2011, sedovaDifferentiableForwardModeling2026, zhengCloseDesigntoManufacturingGap2023, langEfficientStructurePrediction2022, langReviewModelingCure2022}.

By contrast, our design variables parameterize the printing instructions. Then, a differentiable model of the laser-focus PSF, dose accumulation, and polymerization maps this input to the predicted printed structure, on which the electromagnetic objective is evaluated. We also penalize structures that are not fully connected in either the material (which would lead to free-floating structural elements that would be missing after development) or the voids (which would lead to inclusions of liquid photoresist in an otherwise polymerized environment, which are detrimental to functionality). By combining the DLW-Model with structural integrity, true digital blueprints are provided for printable structures with properties on-demand. Finally, these structures can also be made relatively insensitive to variations in writing power during the fabrication process using robust topology optimization methods \cite{sigmundManufacturingTolerantTopology2009}. 

Such a design workflow ensures that the resulting structure is printable, while the optimized printing instructions are precompensated for systematic process-induced deviations, making the entire design process fully fabrication-aware.
We want to emphasize, that by using a dose-accumulation model, the optimizer can make use of fabrication-specific effects such as the proximity effect. By placing single voxels which themselves would not be polymerized, the deposited dose can still affect the polymerization of close-by voxels. The proximity effect is an effect, which cannot be modeled by simple filtering and projection steps, leading to a potentially sub-optimal local minimum when the optimization is not fabrication-aware.
In the following, we introduce the details of such a comprehensive approach to designing photonic structures that are directly feasible for 3D laser-nanoprinting and demonstrate its applicability to several design challenges.
Furthermore, the code for generating the DLW-Model and reproducing most of the presented results is available as open source.

\section{\label{sec:methods}Methods}
One of the most important steps in inverse design is to properly parametrize the problem and convert the parametrization into a fabricable design using a suitable forward model.
Our DLW-Model describes the entire direct-laser-writing process and is illustrated in Fig.~\ref{fig:dlw-scheme} a).
Doing so enables optimization within the experimentally accessible design space and accurate predictions of what the final designs will look like given a writing pattern as input \cite{sedovaDifferentiableForwardModeling2026, zhengCloseDesigntoManufacturingGap2023, langEfficientStructurePrediction2022, langReviewModelingCure2022}.
In general, the density-based topology optimization problem can be formulated as
\begin{align}
     \max\,\,\,&\mathcal{L}(\rho(\mathbf{r}))\\\nonumber
  &\text{s.t.}  \,\,\,\mathbf{c} = \mathbf{0},\\\nonumber
  &\text{s.t.}  \,\,\,\mathbf{d} < \mathbf{0},\\\nonumber
 &\text{s.t.} \,\,\, 0 \le\rho(\mathbf{r}) \le 1\,,\nonumber
\end{align}
where $\mathbf{c}$ represent equality constraints and $\mathbf{d}$ inequality constraints placed on the problem.
For nanophotonics, the density $\rho(\mathbf{r})$ is a representation of the design, which can then be mapped to the relative permittivity $\varepsilon(\rho(\mathbf{r}))$ (or any other physical property) of said design, which is then simulated using a Maxwell solver to calculate the figure of merit $\mathcal{L}(\rho(\mathbf{r}))$. The density will be the starting point in the top left corner of Fig.~\ref{fig:dlw-scheme} a). We describe in the following the design pipeline that cycles clockwise in Fig.~\ref{fig:dlw-scheme} a).

\begin{figure*}[t]
    \centering
    \includegraphics[width=0.9\linewidth]{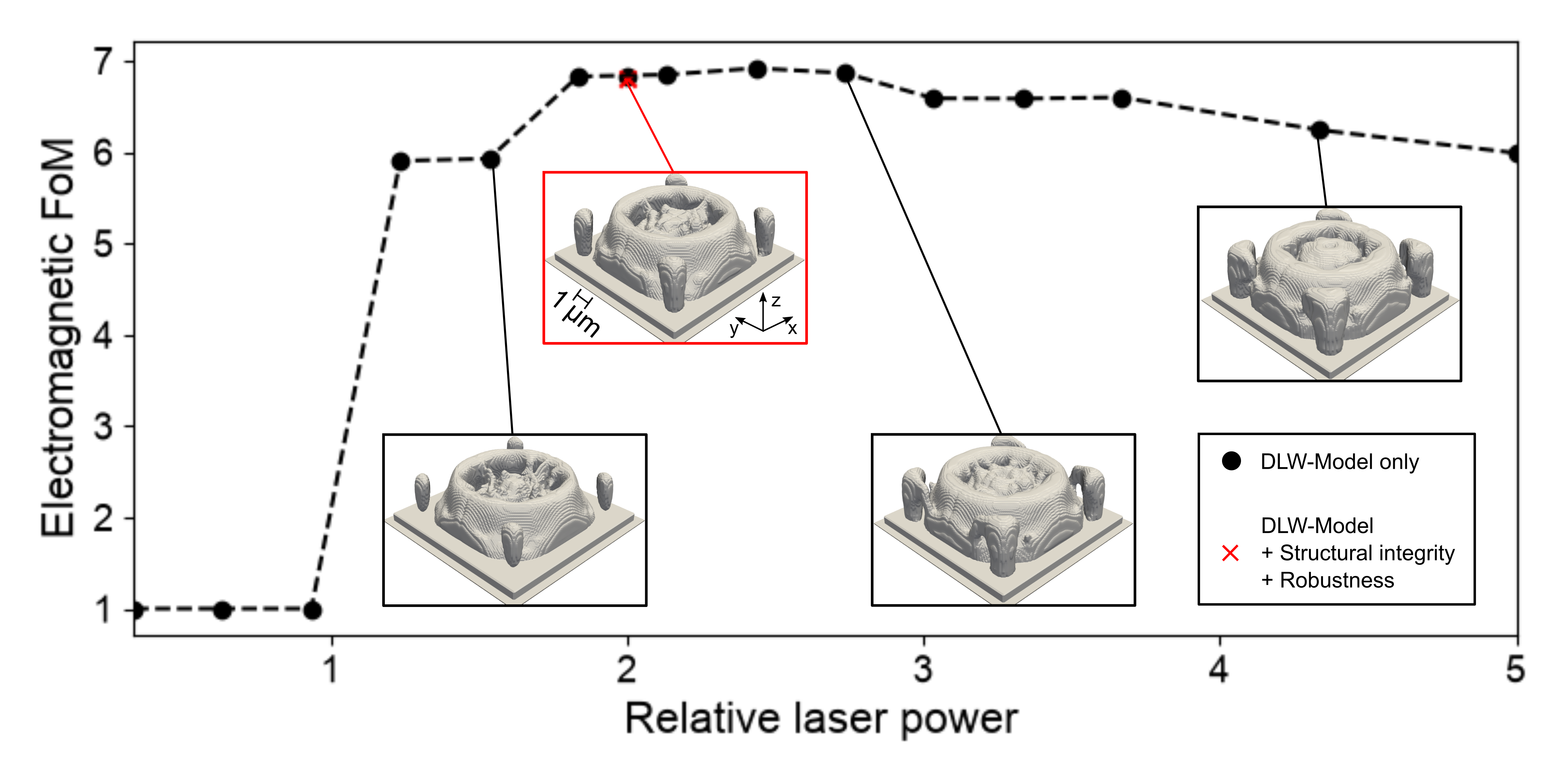}
    \caption{The dependence of the electromagnetic figure of merit ($\mathcal{L}_\text{EM}$) of the small metalens on the relative laser power $P_\text{laser}$. $\mathcal{L}_\text{EM}$ is the measure of how much the light is enhanced at the focal spot compared to no device. We normalized the relative laser power to be $P_\text{laser}=1$ at the polymerization threshold. Each metalens is optimized for a wavelength of $\SI{1.55}{\upmu m}$ and is ${\SI{8}{\upmu m} \times \SI{8}{\upmu m} \times \SI{2.5}{\upmu m}}$ big. For each considered relative laser power, an independent inverse design is performed. Selected metalens designs at different relative laser powers are shown in the insets. A selected device has been designed while additionally considering structural integrity and robustness in the fabrication process ($\SI{15}{\%}$ sensitivity regarding the relative laser power at $P_\text{laser}=2$). The performance of the resulting optimized structure is marked with the red cross. The resulting design is shown in the inset, enclosed by a red bounding box. We observe that incorporating structural integrity and robustness is achievable with only a slight reduction in performance. The red cross marking the electromagnetic performance is only slightly smaller than the corresponding black dot marking the performance of the sample for which integrity and robustness were not enforced.}
    \label{fig:metalens_comparison}
\end{figure*}

\subsection{Writing Pattern}
The input for the 3D laser-nanoprinting process is typically a bit-wise instruction of the laser printing path. 
By controlling where the writing laser is turned on and focused, polymerization in the photoresist can be triggered.
To parametrize the laser path, a writing pattern (\textit{e.g.}, in the form of an .stl file) is used, which can be converted into a set of instructions for the laser path.
Since we are working with density-based topology optimization, we parameterize the bit-wise writing pattern $\sigma(\mathbf{r})$ with the continuous abstract design variable $\rho$, which is approximately binarized in a differentiable manner as

\begin{equation}
    \sigma(\mathbf{r}) = \frac{\tanh(\beta\alpha) + \tanh(\beta(\rho(\mathbf{r}) - \alpha))}{\tanh(\beta\alpha) + \tanh(\beta(1-\alpha))}\,.
    \label{eq:tanh_projection}
\end{equation}
Here, $\alpha$ represents the binarization threshold and $\beta$ the level of binarization, which we also refer to as a binarization step.
As the writing pattern is only used to parametrize the laser path, it does not need to adhere to any fabrication limitations or even be a realistically manufacturable design.
As long as the final, printed design is fabricable, the writing pattern only needs to be a binary voxel distribution in space, giving us more freedom during optimization.

\subsection{Dose Accumulation}
The printed structure is determined by the voxels for which the accumulated exposure dose exceeds the polymerization threshold. We model the accumulated dose as a convolution of the writing pattern $\sigma(\mathbf{r})$ with the squared intensity profile PSF of the laser focus $I_\text{PSF}(\mathbf{r})^2$, thereby accounting for partial exposure of neighboring voxels during the line scan. Therefore, the dose accumulated in the photoresist is given by $D(\mathbf{r})=P_\text{laser}^2\cdot\sigma(\mathbf{r})\ast I^2_\text{PSF}(\mathbf{r})$, where $P_\text{laser}$ represents the relative laser power, which we use as a hyperparameter to control the degree of polymerization for our optimization. Changing the relative laser power will result in larger or smaller features being polymerized, while also accounting for dose accumulation. This also means that control over the relative laser power provides a more accurate description of the fabrication process than simple erosion and dilation alone.

We note that calculating the PSF of the laser focus requires some additional care in the case of 3D laser-nanoprinting and cannot be represented by a simple elongated Gaussian beam. High-NA objective lenses are commonly used for printing when high spatial resolution is required. But simulating a conventional scalar Gaussian beam as the laser focus becomes infeasible at NAs larger than $\sim$0.7, since the paraxial approximation is no longer valid. Therefore, we calculate the laser focus based on vectorial diffraction for an NA 1.4 objective lens, and we account for this in the convolution.
The exact details of how the PSF was modeled numerically are available in the code we provide alongside this publication but it follows approaches described in the literature \cite{richardsElectromagneticDiffractionOptical1959, sedovaDifferentiableForwardModeling2026}.

\subsection{Polymerization}

Using the accumulated dose, we can calculate the degree of polymerization:
\begin{equation}
    \tilde{\sigma}(\mathbf{r}) = \sigma_0\cdot \left(1-e^{-c\cdot D(\mathbf{r})}\right)\, .
\end{equation}
Here, $\sigma_0$ represents the maximum degree of polymerization and $c$ is an experimental prefactor.
By assigning values of $\tilde{\sigma}(\mathbf{r})$ that are above the binarization threshold $\sigma_\text{th}$ as polymerized, we obtain the printed structure $\hat{\tilde{\sigma}}(\mathbf{r})$.
Since the DLW-Model uses a dose accumulation model, voxel-dense areas tend to cross the polymerization threshold more quickly. When using a regular projection such as Eq.~\ref{eq:tanh_projection}, the gradients tend to vanish even at low binarization levels $\beta$ since the designs can become locally binarized early in the optimization, causing the optimization to get stuck in suboptimal local minima.
To alleviate this issue, we employ the Subpixel-Smoothed-Projection (SSP) to binarize the accumulated dose and retrieve the printed structure $\hat{\tilde{\sigma}}(\mathbf{r})$ \cite{hammondUnifyingAcceleratingLevelset2025}.
To simplify the problem for simulation purposes, we assign $P_\text{laser}=1$ as our polymerization threshold and normalize all other parameters of the DLW-Model accordingly.

\subsection{Optical Performance}
After obtaining the final printed structure $\hat{\tilde{\sigma}}(\mathbf{r})$, we need to convert the performance of the device into a scalar figure of merit $\mathcal{L}_\text{EM}(\hat{\tilde{\sigma}}(\mathbf{r}))$ and use its gradients $\nabla\mathcal{L}_\text{EM}(\hat{\tilde{\sigma}}(\mathbf{r}))$ to iteratively converge towards an optimized design.
To obtain the figure of merit, we need to solve Maxwell's equations using the printed design as the permittivity distribution in space.
We use full-wave solvers, such as FDFD (or FDTD as indicated below), to simulate the optical response of our designs.
Our topology optimization problem can then be formulated as
\begin{align}
     \max\,\,\,&\mathcal{L}_\text{EM}(\hat{\tilde{\sigma}}(\mathbf{r}))\\
  &\text{s.t.}  \,\,\, \left(\nabla \times \nabla \times - \omega^2\mu_0\varepsilon_0\varepsilon(\hat{\tilde{\sigma}}(\mathbf{r}))\right)\mathbf{E}(\mathbf{r}, \omega) = -i\omega \mathbf{J}(\mathbf{r}, \omega)\,,\nonumber\\
 &\text{s.t.} \,\,\, 0 \le\hat{\tilde{\sigma}}(\mathbf{r}) \le 1\,,\nonumber
\end{align}
where $\omega$ is the frequency of the light, $\mathbf{E}(\mathbf{r}, \omega)$ is the electric field in the frequency domain, $\mathbf{J}(\mathbf{r}, \omega)$ is a source term, and $\mu_0$ and $\varepsilon_0$ are the vacuum permeability and permittivity respectively. The topology optimization problem can also be formulated equivalently in time domain.

\subsection{Structural Integrity}

The DLW-model ensures that optimization is performed on the structure predicted after printing, rather than on an idealized input geometry. This accounts for fabrication-induced deviations, but it does not by itself guarantee that the printed structure forms a connected, self-supporting object. 
In addition to the structural integrity of the material, it is necessary to ensure that there are no enclosed voids that could trap unpolymerized photoresist and alter the electromagnetic response. 
To promote structural integrity during optimization, we use a virtual temperature method with a nonlinear dependence on both the material and the void \cite{yuIdentificationVoidsInclusions2022, coolPracticalReviewPromoting2025, kusterInverseDesign3D2025}. 
Two sub-objectives are defined for the connectivity of the material $\mathcal{L}_\text{m}(\hat{\tilde{\sigma}}(\mathbf{r}))$ and the void $\mathcal{L}_\text{v}(\hat{\tilde{\sigma}}(\mathbf{r}))$.
We define our full figure of merit as
\begin{align}
    \mathcal{L}(\hat{\tilde{\sigma}}(\mathbf{r})) =&\mathcal{L}_\text{EM}\cdot\left(1-\sqrt{|\text{ReLU}(\mathcal{L}_\text{m}(\hat{\tilde{\sigma}}(\mathbf{r}))|^2 + |\text{ReLU}(\mathcal{L}_\text{v}(\hat{\tilde{\sigma}}(\mathbf{r}))|^2}\right)\,
\end{align}
Full details on how we implement the virtual temperature method are given in Ref.~\cite{kusterInverseDesign3D2025}.

\subsection{Robustness}

As the polymerization threshold can vary with environmental factors, the final printed designs might be over- or underexposed as compared to the optimized designs.
To account for over- and underexposure, we use a robust topology optimization formulation of our problem \cite{sigmundManufacturingTolerantTopology2009}.
We want to highlight that the DLW-Model directly correlates its hyperparameters (\textit{e.g.}, the relative laser power) to the printed design.
Changing the relative laser power will directly lead to an over- or underexposed design with no additional filtering step required.
Typically, the easiest way to handle over- and underexposure (or erosion and dilation in planar topology optimization) is to adjust the binarization threshold. However, without a fabrication-aware forward model, such designs do not necessarily represent an accurate description of fabrication variations.
More accurate methods of modeling over- and underexposure require more complex morphological transformations, which also struggle to capture the influence of the proximity effect on the final designs, making them unsuitable for estimating fabrication variations in a 3D laser-nanoprinting setting.
The DLW-Model eliminates the need to estimate fabrication deviations using heuristic methods and directly provides designs that represent variations in the polymerization threshold (or relative laser power) \cite{svanbergDensityFiltersTopology2013}.

By simulating a design at the desired  relative laser power, an underexposed design at a lower relative laser power, and an overexposed design at a higher relative laser power, we obtain three designs which can be simultaneously optimized.
For the optimization itself, we choose to optimize the worst-performing of the three designs (using the LogSumExp-Function as a differentiable approximation of the minimum function) at each iteration step, making it possible to find designs which perform adequately in all three cases, increasing the tolerance to experimental variabilities.
To characterize the robustness, we use the sensitivity with respect to the relative laser power  ${\Delta P_\text{laser}=\frac{P_\text{target} - P_\text{laser}}{P_\text{laser}}}$, where $P_\text{target}$ is the relative laser power of the under- or overexposed design. For instance, $\Delta P_\text{laser} = \SI{10}{\%}$ at a relative laser power of $P_\text{laser}=2$ would mean that the underexposed design is optimized at $P_\text{laser}=1.8$ while the overexposed design is optimized at $P_\text{laser}=2.2$.

\subsection{Optimization}
Once we compute the final figure of merit $\mathcal{L}(\hat{\tilde{\sigma}}(\mathbf{r}))$ we also need to compute its gradients $\nabla\mathcal{L}(\hat{\tilde{\sigma}}(\mathbf{r}))$ for the optimization. 
To do so, we make sure that every step in our optimization is differentiable by using the adjoint method (for the optical simulation) and automatic differentiation \cite{bradburyJAXComposableTransformations2018, maclaurinAutograd2015, anselPyTorch2Faster2024} for each intermediate step of our simulation. Using the chain rule, we can then propagate all of the individual gradients to compute the full gradient $\nabla\mathcal{L}(\hat{\tilde{\sigma}}(\mathbf{r}))$ in an efficient manner in the so-called backwards simulation.
Finally, we use gradient-based optimizers to update our design variables, iteratively improving our figure of merit until we reach a local minimum which represents our optimized design \cite{johnsonNLopt2017, svanbergMethodMovingAsymptotes1987}.
Further details on the optimization can be found in the supplementary document S1.

\section{Results}
We present several designs which we optimized using the DLW-Model.
First, we present a thorough investigation of how the DLW-Model influences the optimization of a small metalens across different relative laser power levels.
The small metalens is intentionally kept lightweight and can be optimized for lower-end hardware to improve reproducibility.
We then present two additional examples of large-scale topology optimization problems. Specifically, we study a large, high NA metalens and a four-way splitter.

\subsection{Small Metalens}
The small metalens is ${\SI{8}{\upmu m} \times \SI{8}{\upmu m} \times \SI{2.5}{\upmu m}}$ big and placed on top of a substrate. 
Both the substrate and material have a permittivity of $\varepsilon_\text{max} = 1.53^2$, which is chosen to represent typical values found in polymers (such as IP-dip), and are embedded in air $\varepsilon_\text{min} = 1$.
We use a linear interpolation to derive the material from our density $\varepsilon(\hat{\tilde{\sigma}}(\mathbf{r})) = \varepsilon_\text{min} + \hat{\tilde{\sigma}}(\mathbf{r})\cdot(\varepsilon_\text{max} - \varepsilon_\text{min})$.
We illuminate the design with an $x$-polarized plane wave with a wavelength of $\SI{1.55}{\upmu m}$, propagating in the positive $z$-direction. Our design goal is to maximize the $x$-component of the electric field at the focal spot $\SI{1.25}{\upmu m}$ above the design. Simulations of the lens have been made by the FDFD method \cite{luJaxwell2023}.

We define the electromagnetic performance as ${\mathcal{L}_\text{EM} = \frac{|\mathbf{E}_x(\mathbf{r}_0, \omega)|}{|\mathbf{E}_{x,0}(\mathbf{r}_0,\omega)|}}$, where $\mathbf{E}_{x,0}(\mathbf{r}_0, \omega)$ is the value of the $x$-component of the electric field, in the presence of only the substrate, at the focal spot $\mathbf{r}_0$. This means that $\mathcal{L}_\text{EM}$ measures how much the field is enhanced at the focal spot by our design.
Figure~\ref{fig:metalens_comparison} shows multiple small metalenses optimized for various relative laser powers.
Above the polymerization threshold of $P_\text{laser}=1$ relative laser power, we quickly reach highly efficient devices.
Further increasing the relative laser power leads to saturation in the dose accumulation and expands the minimum feature size, ultimately constraining the design space can limit the final performance of the design.
Looking at Fig.~\ref{fig:structures}, we can see that the optimizer can make use of the fabrication-specific effects such as the proximity effect, by placing single, free floating features in the writing pattern. Single voxels which themselves would not be polymerized due to a single exposure can exceed the polymerization threshold due to the exposure of close-by voxels, allowing the optimizer to implement features which would not be obtainable by simple Gaussian filtering.

%We want to emphasize, that by using a dose-accumulation model, the optimizer can make use of fabrication-specific effects such as the proximity effect. By placing single voxels which themselves would not be polymerized, the deposited dose can still affect the polymerization of close-by voxels. The proximity effect is an effect, which cannot be modeled by simple filtering and projection steps, leading to a potentially sub-optimal local minimum when the optimization is not fabrication-aware.
Using the virtual temperature method, the structural integrity of the material and void can be promoted at minimal cost to the performance of the small metalens. The exact numerical details of the implementation, parameter sweeps, resulting designs, and their performance can all be found in the supplementary document S2 and S3.

In the case of the small metalens, the structural integrity and robustness also barely affect the performance of the device, as indicated by the red cross in Fig.~\ref{fig:metalens_comparison}, where a sensitivity with respect to the laser power of $\Delta P_\text{laser}=\SI{15}{\%}$ at $P_\text{laser}=2$ laser power was used.
The final structurally integral, robust design can be seen in Fig.~\ref{fig:structures} a), with the writing pattern on top and the printed design and the electromagnetic field on the bottom.

\subsection{Further Examples}
\begin{figure*}
    \centering
    \includegraphics[width=0.9\linewidth]{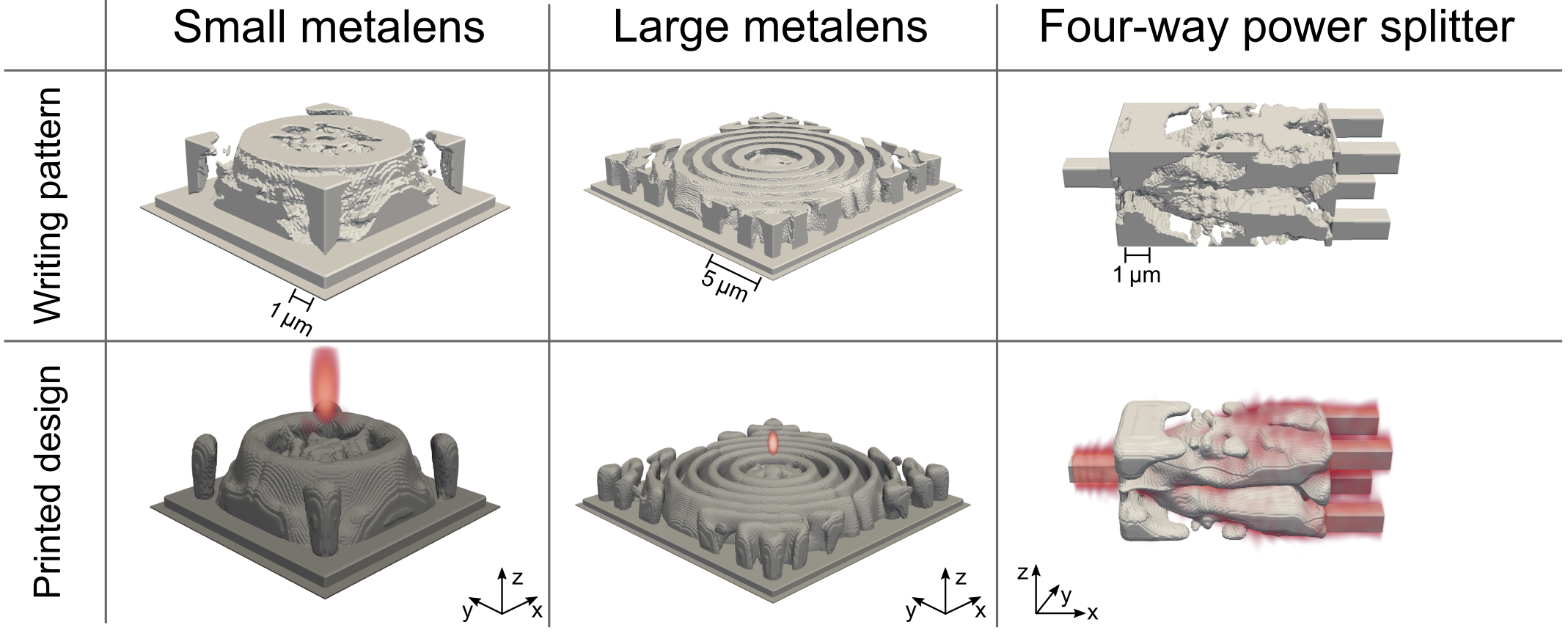}
    \caption{Three different structures optimized using the DLW-Method with structural integrity and robustness. The top figures show the writing patter, whilst the bottom figures show the predicted printed designs with the electric field distribution $|\mathbf{E}|$. a) a small metalens of size ${\SI{8}{\upmu m} \times \SI{8}{\upmu m} \times \SI{2.5}{\upmu m}}$. The small metalens can achieve a field enhancement of up to roughly $6.5$. b) a large metalens of size ${\SI{20}{\upmu m} \times \SI{20}{\upmu m} \times \SI{3}{\upmu m}}$. The large metalens can achieve a field enhancement of roughly $16$ at an NA of over $0.99$. For both metalenses the figure of merit is maximizing the x-component of the incoming electric field at the focal point. c) a four-way power splitter of size ${\SI{10}{\upmu m}\times\SI{5}{\upmu m}\times\SI{5}{\upmu m}}$. The four-way power splitter is able to reach a flux transmission of above $20\%$ into each port. All three examples are optimized for a wavelength of $\SI{1.55}{\upmu m}$. The permittivity of the material is set to $\varepsilon = 1.53^2$ to match the permittivity of typical polymer resists and are placed in air $\varepsilon=1$.
    All printed designs consider structural integrity and robustness.}
    \label{fig:structures}
\end{figure*}

As an example of a large-scale optimization problem, we optimize a large metalens with an NA of almost $1$.
The setup for the large metalens is nearly identical to that of the small metalens, optimized for structural integrity and robustness using the DLW-Model. We merely increase the size of the design region to ${\SI{20}{\upmu m}\times\SI{20}{\upmu m}\times \SI{3}{\upmu m}}$.
The design is optimized at $P_\text{laser}=2$, and we use  $\Delta P_\text{laser}=\SI{10}{\%}$ when optimizing for robustness. We use the FDFD solver again \cite{luJaxwell2023}.
The final design can be seen in Fig.~\ref{fig:structures} b) with the writing pattern at the top and the printed design with the electromagnetic field on the bottom. The final electromagnetic figure of merit of the design is $\mathcal{L}_\text{EM} = 16.1$, with the underexposed design also reaching $\mathcal{L}^\text{under}_\text{EM} = 16.1$. The overexposed design performs slightly worse with $\mathcal{L}^\text{over}_\text{EM} = 15.2$.

Our third example is a medium-sized optimization problem in which we design a four-way power splitter that makes full use of the third dimension unlocked by 3D nanoprinting, also optimized with structural integrity and robustness.
The four-way power splitter design uses one input waveguide and four output waveguides, with the design region connecting the input and output waveguides. The optimization goal is to evenly split the incoming light between the output waveguides, ideally with a total flux of $\SI{25}{\%}$ of the input power exiting each waveguide.
The design region has a size of ${\SI{10}{\upmu m}\times\SI{5}{\upmu m}\times\SI{5}{\upmu m}}$.
We use symmetry to reduce our effective design region down to ${\SI{10}{\upmu m}\times\SI{2.5}{\upmu m}\times\SI{2.5}{\upmu m}}$.
The waveguides are chosen to be quadratic with a width of $\SI{1}{\upmu m}$. The input waveguide is placed at the center of the design region, while the output waveguides are placed halfway to the middle of the design region in the $y$- and $z$-direction.
The permittivity of the design is set to represent a polymer with $\varepsilon = 1.53^2$, and the design is embedded in air with $\varepsilon = 1$.
We excite the fundamental mode of the input waveguide at a vacuum wavelength of $\lambda_0 = \SI{1.55}{\upmu m}$, propagating in the $x$-direction.
Our electromagnetic figure of merit is then defined as the coupling efficiency of the fundamental mode into one output waveguide
\begin{equation}
    \mathcal{L}_\text{EM} = \text{CE}_\text{TE0}\, .
\end{equation}
We use the FDTD solver Tidy3D to simulate the problem. 
The final structure can be seen in Fig.~\ref{fig:structures} c) with the writing pattern on top and the printed design with the electromagnetic field on the bottom. 
Each output waveguide can reach a flux transmission of $T_\text{flux}=\SI{21}{\%}$, resulting in a total coupling efficiency of over $\SI{80}{\%}$ for the design. The overexposed design reaches $T^\text{over}_\text{flux} = \SI{19}{\%}$ for a single waveguide, while the underexposed design reaches a similar flux transmission as the regular four-way power splitter at $T^\text{under}_\text{EM}=\SI{21}{\%}$.

\section{Conclusion}
We present a fabrication-aware design pipeline for the inverse design of 3D laser-nanoprinted devices. 
By replacing the traditional filtering and projection approach used in topology optimization in nanophotonics with a forward model of the 3D laser-nanoprinting process, we obtain optimized structures within a design space that fully respects the limitations of the fabrication method.
The DLW-Model includes descriptions of the PSF and dose accumulation in the photoresist, enabling us to deliver precompensated design files optimized for the performance of the (simulated) printed structure.
The writing pattern can be converted into a suitable file format for the 3D nanoprinting machine and is then converted into the laser-writing path.
To account for additional experimental constraints, we also include a virtual temperature model to ensure the structural connectivity of the printed material and the remaining voids. 
By adding robustness to over- or underexposed designs, we can also directly accommodate variations in experimental parameters outside of our control, which might influence the polymerization threshold of the photoresist.

\begin{backmatter}
\bmsection{Funding}
O. K., C. R., and M. W. acknowledge support through the Deutsche Forschungsgemeinschaft (DFG, German Research Foundation) under Germany’s Excellence Strategy via the Excellence Cluster 3D Matter Made to Order (EXC-2082/2, Grant No. 390761711) and from the Carl Zeiss Foundation via CZF-Focus@HEiKA. C. R. and T. J. S. acknowledge support
by the Helmholtz Association via the Helmholtz program “Materials Systems Engineering” (MSE). 

\bmsection{Acknowledgments}
The authors gratefully acknowledge the computing time provided on the high-performance computer HoreKa by the National High-Performance Computing Center at KIT (NHR@KIT). This center is jointly supported by the Federal Ministry of Education and Research and the Ministry of Science, Research and the Arts of Baden-Württemberg, as part of the National High-Performance Computing (NHR) joint funding program (https://www.nhr-verein.de/en/our-partners). HoreKa is partly funded by the German Research Foundation (DFG).

We thank Flexcompute for providing a license for Tidy3D, which was used for the electromagnetic simulations in this work.

\bmsection{Data Availability Statement}
Data underlying the results presented in this paper are not publicly available but may be obtained from the authors upon reasonable request or generated from the publicly available code under \url{https://github.com/OlloKuster/direct_laser_writing}.

\bmsection{Supplementary Document}
We provide supplementary document alongside this publication. The supplementary document includes a section describing details on the optimization procedure, a section on the numerical details of the structures and a section showing parameter sweeps regarding the minimum feature size, relative laser power, connectivity constraints and sensitivity with respect to the laser power.

\end{backmatter}

% Bibliography
\bibliography{article}

% Full bibliography added automatically for Optics Letters submissions; the following line will simply be ignored if submitting to other journals.
% Note that this extra page will not count against page length
\bibliographyfullrefs{article}

%Manual citation list
%\begin{thebibliography}{1}
%\bibitem{Zhang:14}
%Y.~Zhang, S.~Qiao, L.~Sun, Q.~W. Shi, W.~Huang, %L.~Li, and Z.~Yang,
 % \enquote{Photoinduced active terahertz metamaterials with nanostructured
  %vanadium dioxide film deposited by sol-gel method,} Opt. Express \textbf{22},
  %11070--11078 (2014).
%\end{thebibliography}

\section{Supplementary Document}
\subsection{Optimization of the small metalens and further examples}
We have found that optimizing the small metalens yields better results when using a free-form optimization without the DLW-Model as an initial guess. By optimizing a small metalens without any minimum feature size constraints for $15$ iterations at binarization steps $\beta=16$ and $\beta=32$ each and then using the resulting design as an input for the optimization using the DLW-Model as described at full binarization $\beta=\infty$, we are able to converge to better performing devices.

Each device is initialized at $\rho = 0.5$ everywhere. We note that, in general, randomized initialization can be challenging for the DLW-Model, as dose accumulation typically converges to $\hat{\tilde{\sigma}}=1$ almost everywhere.

\begin{figure}
    \centering
    \includegraphics[width=\linewidth]{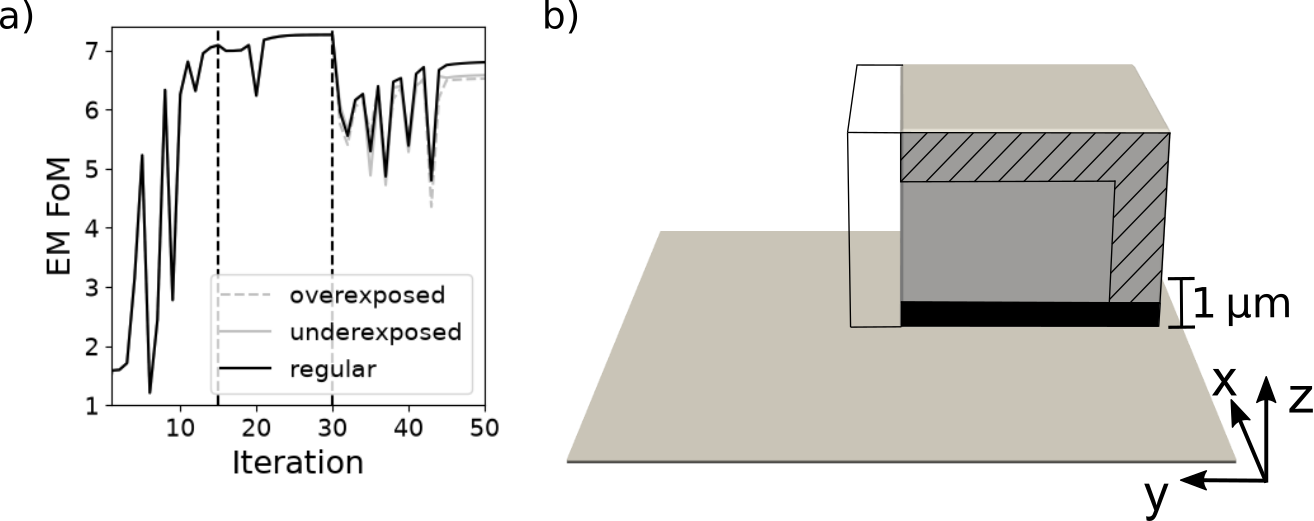}
    \caption{a) The loss function of a full optimization for a small metalens, including structural integrity and robustness. Each dashed vertical line represents an increase in the binarization level starting from $\beta=16$ to $\beta=32$ and finally $\beta=\infty$ using the SSP projection. For the first $30$ iterations, the optimization was performed in a free-form manner, using only the SSP projection and no minimum-feature-size filter. After $30$ iterations, the DLW-Model, virtual temperature method, and robustness are included. The performance of the overexposed and underexposed designs is also indicated.
    b) Sketch of the computational setup of the small metalens. The dashed region indicates the buffer region to avoid sharp cutoffs after filtering. The solid black region indicates where material was enforced to prevent the structure from detaching from the substrate after filtering and to increase binarization during the optimization. We also indicate the extension of the design region with a wiregrid to avoid filtering effects at the symmetry planes of the design.}
    \label{fig:a_setup}
\end{figure}

An example optimization of the small metalens is shown in Fig.~\ref{fig:a_setup} a), where a free-form optimization was performed for the first $30$ iterations. Afterward, we include structural integrity, robustness, and the DLW-Model in the optimization for another $20$ steps.

We use the Method of Moving Asymptotes (MMA) for all of our optimizations.
Each optimization step for the small metalens takes roughly one minute on an NVIDIA RTX A4500 GPU. The entire optimization of the small metalens is performed in an open-source, lightweight manner to ensure reproducibility.

The large metalens is optimized using an NVIDIA A100 GPU and takes roughly $\SI{20}{\text{min}}$ per iteration (which includes three simulations for robustness).

The four-way power splitter using Tidy3D takes less than $\SI{10}{\text{min}}$ per iteration (which includes three simulations for robustness).

\subsection{Numerical details of the implementation}
Convolution filters tend to run into the issue that sharp cutoffs might appear at the edges of the design region, which cannot be fabricated.
To avoid this issue, we implement a $\SI{1}{\upmu m}$ buffer zone that extends the design region at the relevant surfaces, allowing the design to extrude into the buffer zones.
We then apply a mask that forces the design region to be $0$ in the relevant regions before filtering the design.
Furthermore, for the metalens, we also enforce a $\SI{0.5}{\upmu m}$ material layer at the bottom of the design region to improve connectivity as the binarization level increases, which otherwise might detach the bottom of the design from the substrate.
Similarly, we extend the waveguides into the design region of the four-way power splitter.
A sketch of the setup with the described buffers and masks is displayed in Fig.~\ref{fig:a_setup} b).

\subsection{Small Metalens designs for various combinations of parameters}
\begin{figure*}
    \centering
    \includegraphics[width=0.75\linewidth]{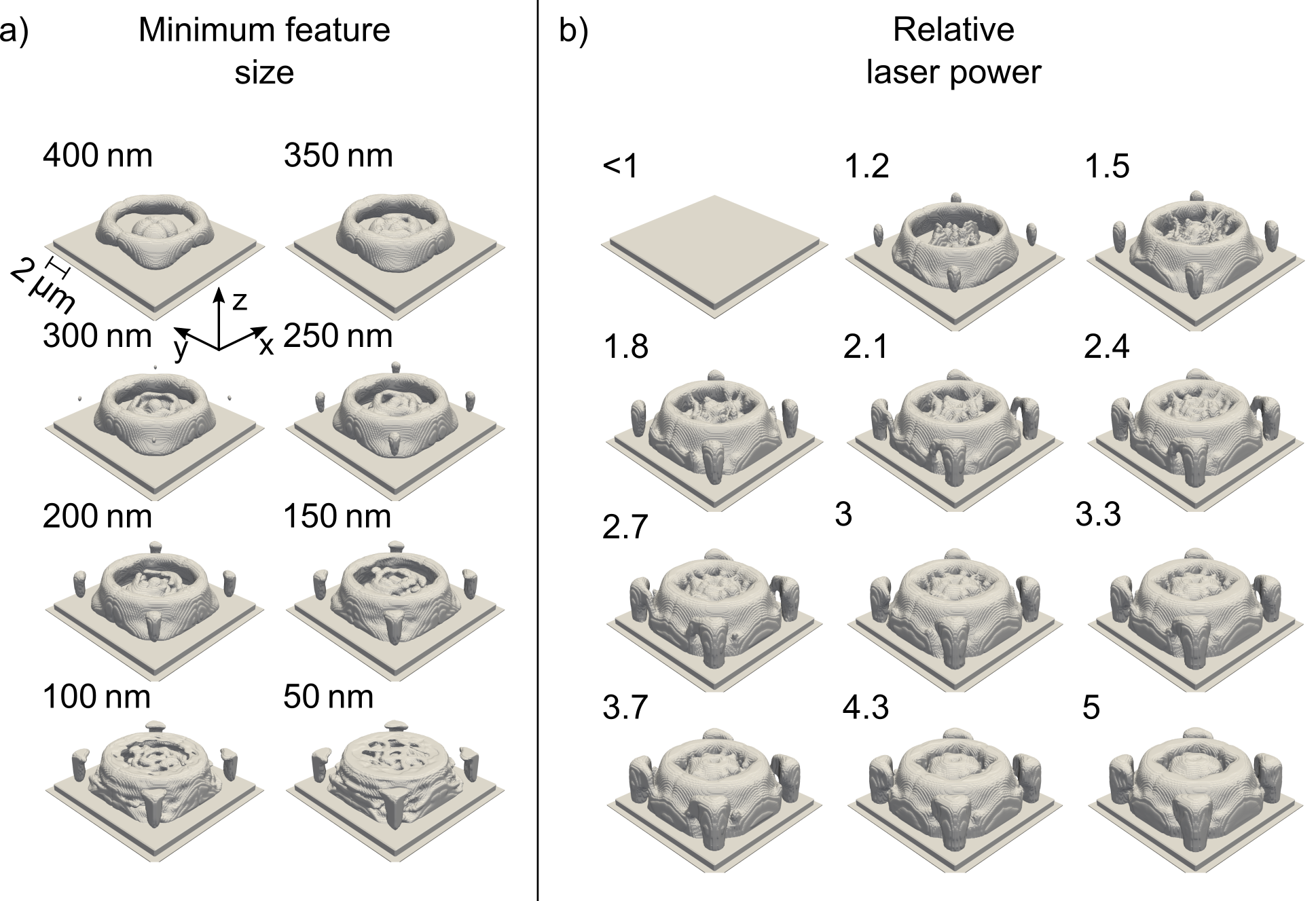}
    \caption{a) The optimized metalens designs using regular filtering and projection depending on the enforced minimum feature size. b) The optimized metalens designs using the DLW-Model depending on the relative laser power.}
    \label{fig:a_feature_size}
\end{figure*}

\begin{figure*}
    \centering
    \includegraphics[width=0.75\linewidth]{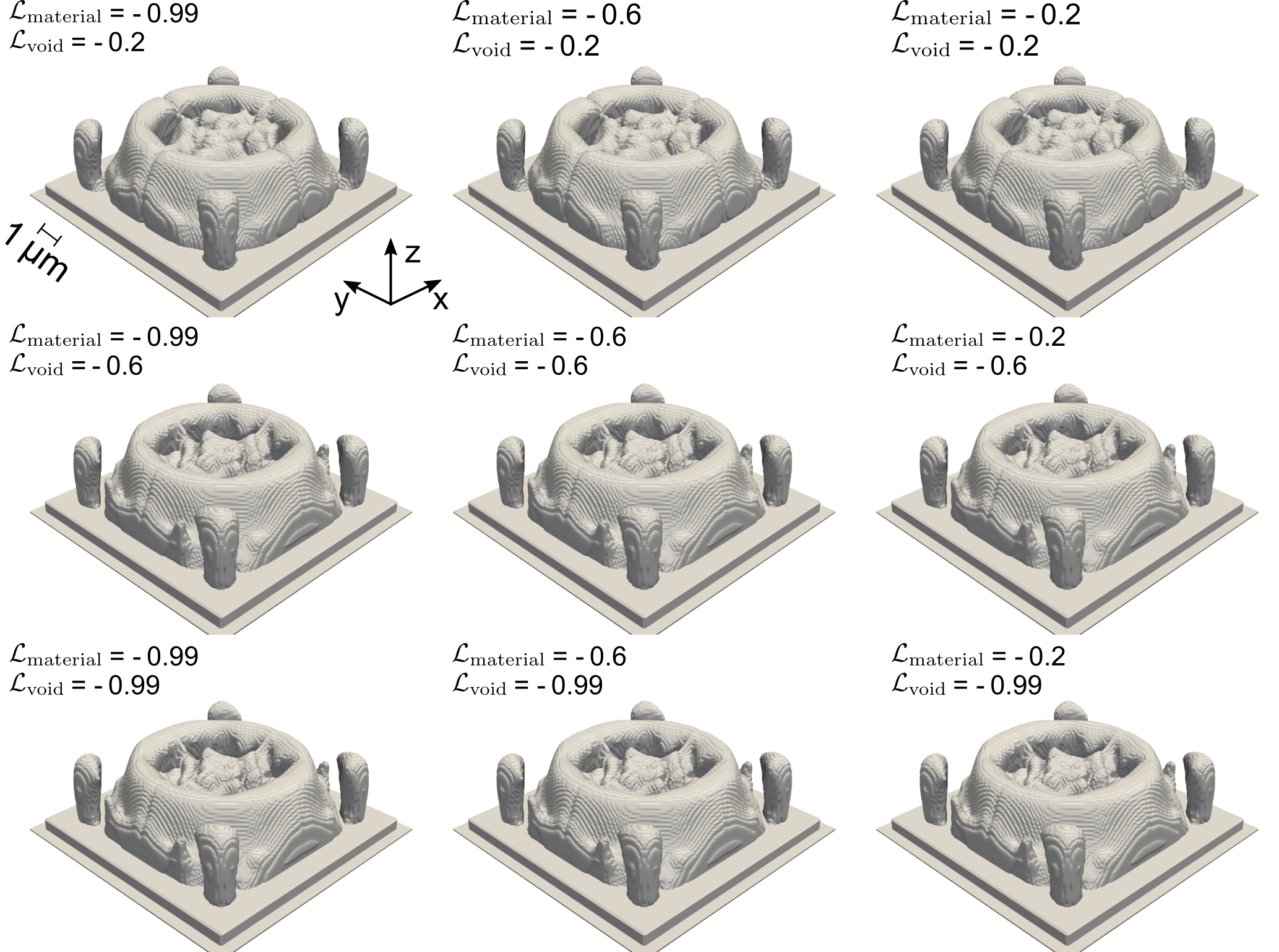}
    \caption{The optimized metalens designs using the DLW-Model and with enforced structural integrity. The values for $\mathcal{L}_\text{material}$ and $\mathcal{L}_\text{void}$ denote the initial values used at the start of the optimization and correspond to a density of $\rho = 0.5$ everywhere. These initial values are then used to derive the threshold values $\tau_\text{material}$ and $\tau_\text{void}$.} 
    \label{fig:a_heat}
\end{figure*}

\begin{figure*}
    \centering
    \includegraphics[width=0.75\linewidth]{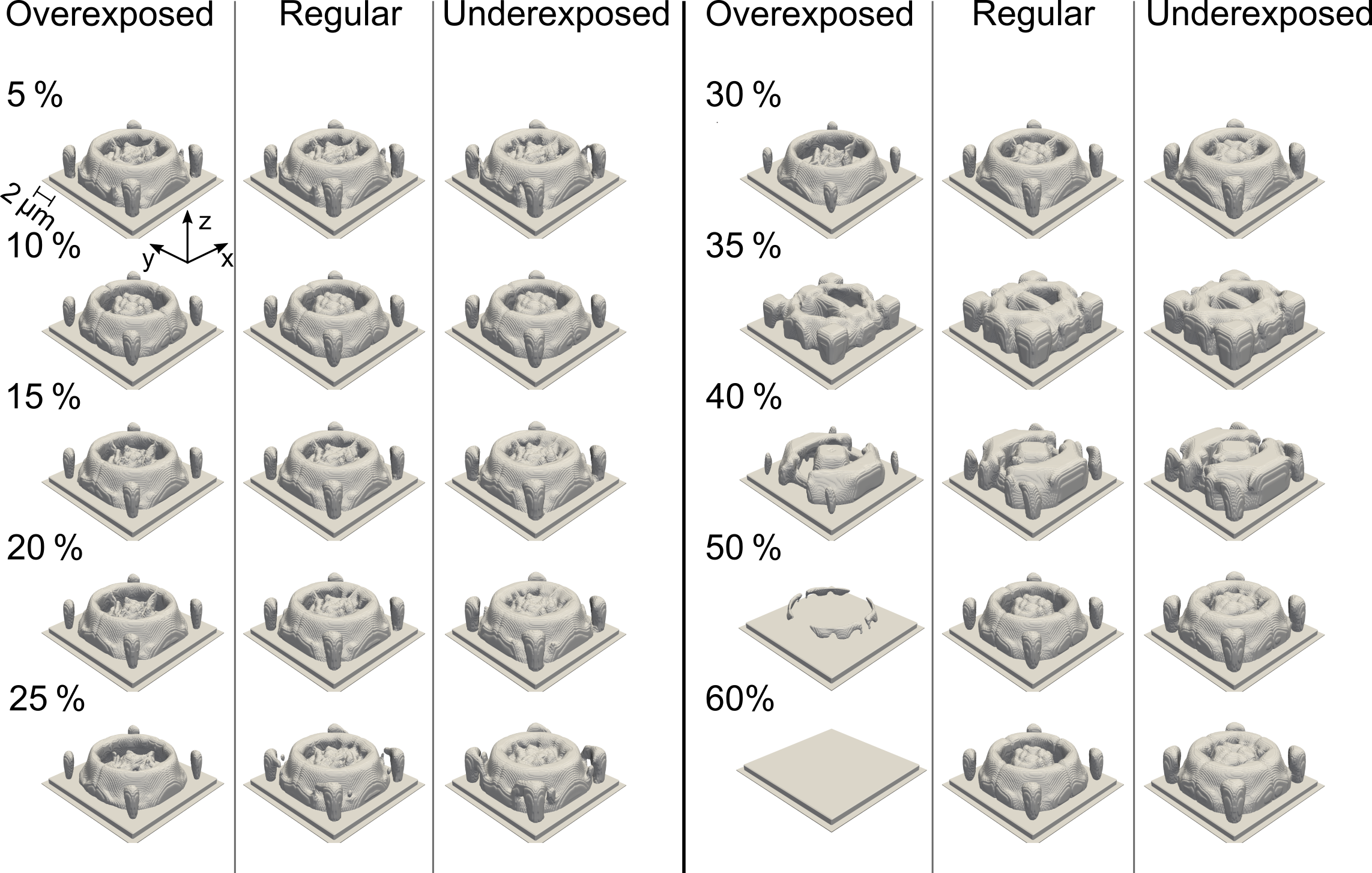}
    \caption{The optimized metalens designs using the DLW-Model, with structural integrity and robustness added. The designs are optimized at a relative laser power of $P_\text{laser}=2$, and we sweep over the sensitivity with respect to the laser power at $P_\text{laser}=2$.}
    \label{fig:a_robust}
\end{figure*}

To determine the best-performing devices within the available parameter space for the small metalens, multiple parameter sweeps have been conducted.
The resulting structures are shown in Fig.~\ref{fig:a_feature_size} for the minimum feature size, in Fig.~\ref{fig:a_heat} for the structural integrity, and in Fig.~\ref{fig:a_robust} for robustness.


\end{document}